\documentclass[12pt]{article}
\pdfoutput=1
\usepackage[top=30truemm,bottom=30truemm,left=25truemm,right=25truemm]{geometry}
\usepackage{authblk}
\usepackage{amsthm}
\usepackage{amsmath,amssymb,bm,mathtools}
\usepackage{cite}
\usepackage[]{hyperref}
\hypersetup{colorlinks,linkcolor={blue},citecolor={blue},urlcolor={blue}}  
\usepackage[titletoc]{appendix}
\usepackage{url}
\usepackage{graphicx,color}
\usepackage{braket}
\usepackage{qcircuit}
\usepackage{here}
\usepackage{mathrsfs}
\usepackage{subcaption}
\usepackage{float}

\usepackage{titlesec}

\titleformat*{\section}{\large\bfseries}

\newtheorem*{definition*}{Definition}
\newtheorem*{assumption*}{Assumption}

\def\sinh{{\mathrm{sinh}}}
\def\cosh{{\mathrm{cosh}}}

\def\be{\begin{equation}}
\def\ee{\end{equation}}
\def\ba{\begin{eqnarray}}
\def\ea{\end{eqnarray}}

\numberwithin{equation}{section}

\AtBeginDocument{%
  {\footnotesize\global\footnotesep=0.8\baselineskip}%
}

\begin{document}

\begin{titlepage}
\thispagestyle{empty}

\begin{flushright}
\end{flushright}

\bigskip

\begin{center}
\noindent{\bf \Large Crosscap Quenches and Entanglement Evolution}\\

\vspace{1.4cm}

{\bf Zixia Wei$^{a}$ and Yasushi Yoneta$^{b}$}
\vspace{1cm}\\

{\it 
$^a$Jefferson Physical Laboratory, 
Harvard University, Cambridge, MA 02138, USA
}\\[1.5mm]


{\it
$^b$Center for Quantum Computing, RIKEN, Wako, Saitama 351-0198, Japan
}\\[1.5mm]

\vskip 3.5cm
\end{center}

\begin{abstract}
Understanding the mechanisms by which complex correlations emerge through the dynamics of quantum many-body systems remains a fundamental challenge in modern physics. To address this, quench dynamics starting from nonthermal states have been extensively studied, leading to significant progress. In this paper, we propose a novel quench protocol, termed the ``crosscap quench'', to investigate how highly structured thermal pure states relax into typical ones. We begin by analyzing conformal field theories (CFTs) and derive universal features in the time evolution of the entanglement entropy. Furthermore, leveraging the AdS/CFT correspondence, we study holographic CFTs, providing an analytically tractable example in chaotic CFTs. Finally, we validate these findings through numerical simulations in both nonintegrable and integrable quantum spin systems.
\end{abstract}

\end{titlepage}

\newpage
\setcounter{page}{1}
\tableofcontents


\section{Introduction}

Understanding the dynamics of quantum many-systems is one of the most important challenges in statistical physics, and quantum quenches \cite{CC05,CC06,CC16} provide a rich playground for this purpose. A quantum quench is a unitary evolution starting from a pure initial state, triggered by a sudden change of parameters in the system. Through the quench dynamics, we can see how a highly organized initial state evolves into thermal equilibrium in chaotic systems. 

In this paper, instead of focusing on the thermalization dynamics, we study a new type of quench where the initial state is already in thermal equilibrium. The initial states we would like to choose are the entangled antipodal pair (EAP) states \cite{CY24,Yoneta24}. 
Consider a quantum spin-$1/2$ system on a circle with length $2N$. The EAP state is defined as 
\begin{align}
    \ket{\rm EAP} \equiv \bigotimes_{j=1}^{N} \ket{\Phi}_{j,j+N}, \label{eq:EAP}
\end{align}
where $\ket{\Phi}_{j,j+N}$ is a maximally entangled state between antipodal pairs of spins at site $j$ and $j+N$. For an arbitrary system, the EAP states are in the microscopic thermal equilibrium (MITE) \cite{MIKU17} at infinite temperature, in the sense that they are locally indistinguishable from the canonical Gibbs state with the same temperature, since the reduced density matrix for any geometrically local subsystem $A$ with length smaller than or equal to $N$ is the maximally mixed state. Furthermore, for a 1D Hamiltonian $H$ possessing a certain antiunitary symmetry, such as time-reversal or complex conjugate symmetry, the imaginary-time evolved EAP state
\begin{align}\label{eq:EAP_cutoff}
    \ket{\rm EAP(\beta)} \equiv \frac{e^{-\frac{\beta}{4}H}\ket{\rm EAP}}{\sqrt{\braket{{\rm EAP}|e^{-\frac{\beta}{2}H}|{\rm EAP}}}}
\end{align}
serves as a thermal pure state at inverse temperature $\beta$.
Here, by a thermal pure state at inverse temperature, we mean a pure state whose expectation value of any geometrically local observable agrees with that of the Gibbs state at the same temperature, up to corrections that vanish in the thermodynamic limit.
Thanks to it highly organized entanglement structure, $\ket{\rm EAP(\beta)}$ can be efficiently prepared using the matrix product states (MPS) \cite{Yoneta24}, despite this, it exhibits a volume law entanglement.

While $\ket{\rm EAP(\beta)}$ serves as a thermal pure state, it is a highly atypical one. In fact, it can be easily distinguished from {a typical state by} the antipodal correlation functions, {which are defined on a few degrees of freedom but are considered highly nonlocal in 1D systems. In this sense, the EAP state utilizes the antipodal entanglement structure to deceive local observers into perceiving thermal equilibrium and at the same time stays easy to simulate.}

As a result, even though $\ket{\rm EAP(\beta)}$ is already in thermal equilibrium, its correlation structure will get more scrambled under chaotic quenches.
We will study the time evolution of the entanglement entropy after such quenches. In contrast to correlation functions, the entanglement entropy is defined independently of the choice of fields, making it a useful tool for highlighting a hierarchy between different notions of thermal equilibrium. In particular, EAP states exhibit a remarkable property in which a single interval is maximally entangled with its complement, whereas an antipodally located double interval is completely separable from its complement. We will investigate how this sharp entanglement structure evolves toward that of a typical state. We will also study time evolution of EAP states in integrable systems to elucidate the role of nonintegrability of the system.

Among all the variations, quantum quenches in conformal field theories (CFT) are one of the most well-studied class \cite{CC16} which is under good analytic control. Therefore, we would like to study quenches from EAP states in both lattice systems and CFTs. 

In order to formulate EAP quenches in CFTs, it is useful to note that EAP states in lattice systems are profoundly related to crosscap states in 2D CFT. In 2D CFT, a crosscap state is a state living on the Hilbert space supported on a circle that is realized by performing the Euclidean path integral over a ``crosscap'', a spacetime structure obtained by cutting a hole and identifying its antipodal points on a 2D Euclidean manifold. In \cite{CK21}, the authors extend the notions of crosscap states to a class of integrable field theories and a class of integrable spin chains, where those in integrable spin chains have the same form as the EAP states in Eq.~\eqref{eq:EAP}. 

While it remains unclear what the exact relation between crosscap states is in {\it generic} CFTs and EAP states in spin systems \cite{Tan2024}, we will compute some universal behavior of the entanglement entropy in CFT crosscap states and argue that they essentially capture the entanglement structure exhibited by the EAP states. Based on this, in the CFT cases, we will study quantum quenches starting from the crosscap states, which we refer to as ``crosscap quenches'' in this paper.

In section \ref{sec:CQinCFT}, we first formulate the crosscap quench in CFT, and use the replica trick to study some universal behaviors of entanglement R\'enyi entropy in the initial state to see that it essentially captures the entanglement structure in EAP states. 
After that, we will use the words ``crosscap state (quench)'' and ``EAP state (quench)'' interchangeably. 
Then we will study some universal behaviors of time evolution of EE after the crosscap quench in CFTs. In section \ref{sec:HolographiCQ}, we identify the gravity dual of the crosscap quench in AdS$_3$/CFT$_2$ and study the holographic entanglement entropy. This serves as an analytically tractable example of a crosscap quench in a chaotic CFT. In section \ref{sec:chaotic_spin_system}, we study the crosscap quench in nonintegrable spin systems, and confirm its qualitative consistency with the holographic results. In section \ref{sec:integrable_spin_system}, we study the crosscap quench in integrable spin systems, and contrast its differences from those in chaotic systems. 

{\bf Note added:} While this paper is under preparation, a very interesting paper \cite{CCR24} came out and also studied time evolution of the entanglement entropy from crosscap states but in different systems. The authors of \cite{CCR24} focused on quantum circuit models and integrable fermion chains while we focus on CFT, holography and spin chains in this paper. The integrable fermion chains investigated in \cite{CCR24} should be equivalent to our computations for the transverse-field Ising models in section \ref{sec:integrable_spin_system} via the Jordan-Wigner transformation in some regimes.

\section{Crosscap Quench in CFT and Entanglement R\'enyi Entropy}\label{sec:CQinCFT}

In this section, we first formulate the crosscap quench in CFT and explain how the entanglement R\'enyi entropy can be computed using the replica trick. We then study some universal behaviors of the initial state to establish a connection between the crosscap states and the EAP states. After that, we study some universal behaviors appearing in the time evolution. 

Let us start by thinking about a Euclidean path integral over a cylinder whose spatial direction is a circle $S^1$ parameterized by $0\leq x < 2L$ with $x \sim x+2L$, and Euclidean time direction is an interval parameterized by $-\alpha \leq \tau \leq 0$. Imposing the boundary condition $\ket{\psi}$ at $\tau = -\alpha$ and $\bra{\varphi}$ at $\tau = 0$, the path integral over the cylinder computes $\braket{\varphi|e^{-\alpha H}|\psi}$. Therefore, by changing the boundary condition $\bra{\varphi}$ at $\tau = 0$, the path integral realizes the wave functional of state $e^{-\alpha H}|\psi\rangle$ at $\tau = 0$. 

Now, instead of explicitly imposing a boundary condition $\ket{\psi}$ at $\tau = -\alpha$, we glue it by performing an antipodal identification $x \sim x+L$ at $\tau = -\alpha$. Such a structure is called a crosscap. Due to the crosscap, now the Euclidean path integral only has one boundary at $\tau = 0$ and the topology becomes a M\"obius strip. Performing the Euclidean path integral over the crosscap is equivalent to imposing a crosscap state $\ket{C}$ as the ``boundary'' condition of $\tau = -\alpha$. The crosscap state satisfies the following constraints from consistency condition of 2D CFT \cite{Ishibashi88}
\begin{align}\label{eq:crosscap_constraint}
    (L_n - (-1)^n \bar{L}_{-n}) \ket{C} = 0,~~~(n = 0,1,2,\cdots), 
\end{align}
where $L_n$ ($\bar{L}_{n}$) are holomorphic (anti-holomorphic) Virasoro generators. 

\begin{figure}[h]
    \centering
    \includegraphics[width=14cm]{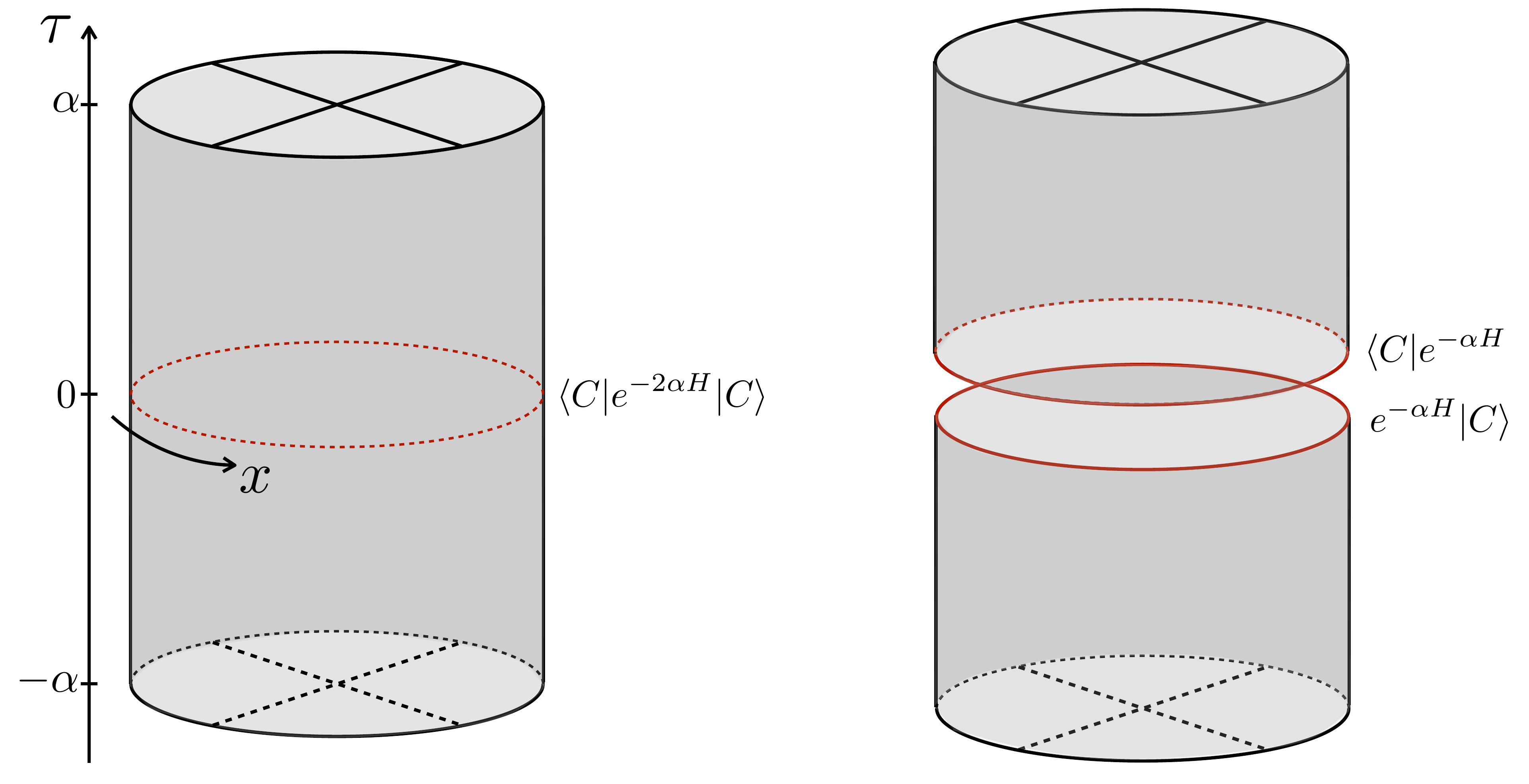}
    \caption{The left panel shows a Klein bottle which is parameterized by $-\alpha \leq \tau \leq \alpha $ and $0\leq x < 2L$ with $x \sim x+2L$. At $\tau = \pm \alpha$, there are two crosscaps, i.e. the antipodal points are identified as $x\sim x + L$.
    The CFT partition function on such a Klein bottle computes $\bra{C}e^{-2 \alpha H}\ket{C}$. Cutting the Klein bottle into two at $\tau = 0$, the lower half corresponds to $e^{-\alpha H}\ket{C}$ and the upper half corresponds to $\bra{C}e^{-\alpha H}$, as shown in the right panel. }
    \label{fig:Klein_bottle}
\end{figure}

As a result, the path integral over such a M\"obius strip realizes $e^{-\alpha H}|C\rangle$ on its boundary $\tau = 0$. Accordingly, $\bra{C}e^{-\alpha H}$ is realized by its time reflection, and the densiy martrix $e^{-\alpha H}|C\rangle\langle C|e^{-\alpha H}$ is realized by a combination of them. 
The inner product $\langle C|e^{-2\alpha H}|C\rangle$ is given by gluing the two boundaries together, and as a result, it is given by a path integral over a tube with two crosscaps on the two ends, which is equivalent to a Klein bottle $\mathbb{K}^2$,
\begin{align}
    \langle C|e^{-2\alpha H}|C\rangle = Z_{\mathbb{K}^2},
\end{align}
as sketched in Fig.~\ref{fig:Klein_bottle}.
Let us then take the interval $0\leq x \leq l$ as the subsystem $A$, and use $A^C$ to denote its complement. The (unnormalized) reduced density matrix $\rho_A \equiv {\rm Tr}_{A^C} \left(e^{-\alpha H}|C\rangle\langle C|e^{-\alpha H}\right)$ on $A$ can be realized by gluing the $A^C$ part of the bra M\"obius strip and the ket M\"obius strip,
leaving the boundaries corresponding to $A$ open. 

\begin{figure}
    \centering
    \includegraphics[width=14cm]{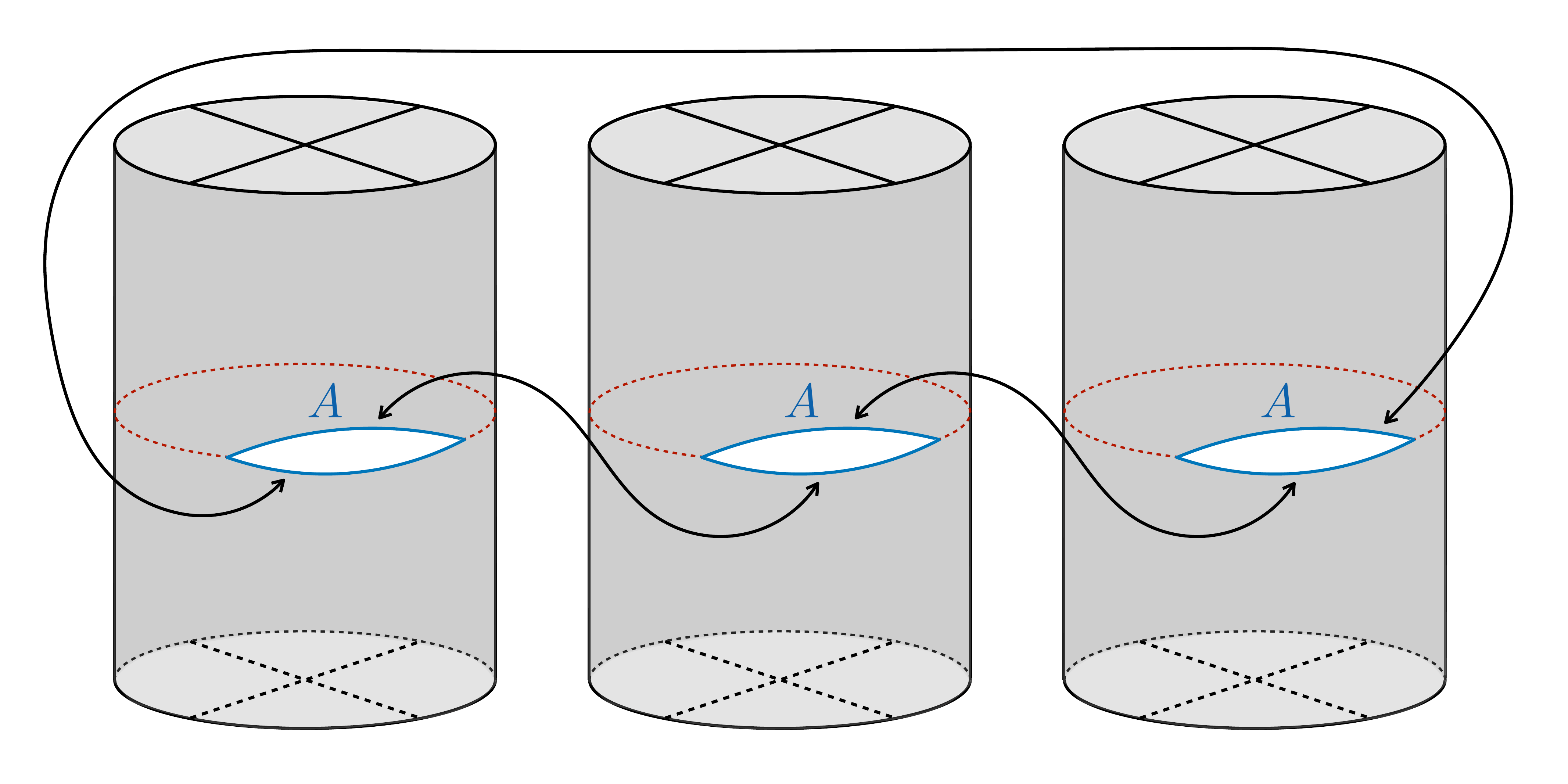}
    \caption{The 3-replicated manifold $\Sigma_3$ corresponding to $\rho^3_A$. }
    \label{fig:3replica}
\end{figure}

Based on this, we can apply the replica method developed in \cite{CC04} to compute the entanglement R\'enyi entropy of subsystem $A$. We can realize $\rho_A^n$ by preparing $n$ copies of $\rho_A$, and then gluing the upper half of the slit in the $i$-th copy to the lower half in the $(i+1)$-th copy for $1\leq i\leq n-1$,
and ${\rm Tr}\left(\rho_A^n\right)$ by gluing the two remaining two boundaries in $\rho_A$. See Fig.~\ref{fig:3replica} for a sketch of the $n=3$ case.
This is a path integral over a $n$-replicated manifold $\Sigma_n$ with $2n$ crosscaps. 
\begin{align}
    {\rm Tr}\left(\rho_A^n\right) =Z_{\Sigma_n}.
\end{align}
With this notation, $\Sigma_1 = \mathbb{K}^2$. 
As a result, the entanglement R\'enyi entropy of the subsystem $A$ can be computed as 
\begin{align}
    S^{(n)}_A = \frac{1}{1-n} \log \frac{{\rm Tr}\left(\rho_A^n\right)}{\left({\rm Tr}\rho_A\right)^n} = \frac{1}{1-n} \log \frac{Z_{\Sigma_n}}{\left(Z_{\Sigma_1}\right)^n}. 
\end{align}
The explicit evaluation of this expression requires computing the partition function on the replicated manifold $\Sigma_n$, which is often complicated. Instead of thinking of $Z_{\Sigma_n}$ as the partition function of $n$-replicated manifolds, it turns out to be useful to regard it alternatively as a $n$-replicated theory on the original manifold $\Sigma_1$, with one twist operator $\sigma_n$ inserted at $(\tau,x) = (0,0)$, and another twist operator $\tilde{\sigma}_n$ inserted at $(\tau,x) = (0,l)$ \cite{CCD07}. Given a local operator $O^{(i)}$ in the $i$-th copy of the theory and moving it around $\sigma_n$ ($\tilde{\sigma}_n$) with angle $2\pi$, it turns into the corresponding local operator $O^{(i+1)}$ ($O^{(i-1)}$) in the $(i+1)$-th ($(i-1)$-th) copy of the theory. The twist operators behave as primary operators with chiral and anti-chiral conformal dimension 
\begin{align}
    h_n = \bar{h}_n = \frac{c}{24} \left(n-\frac{1}{n}\right). 
\end{align}
In this description, the entanglement R\'enyi entropy can be computed with the correlation function of the twist operators on the Klein bottle \cite{CCD07}, 
\begin{align}
    S^{(n)}_A = \frac{1}{1-n} \log \braket{\sigma_n(0,0)\tilde{\sigma}_n(0,l)}_{\mathbb{K}^2}, 
\end{align}
where the correlation function is computed in the replicated theory, whose central charge is $nc$. In general, if the subsystem $A$ is taken as a multi-interval between $(\tau_{a_i},x_{a_i})$ and $(\tau_{b_i},x_{b_i})$, the corresponding entanglement R\'enyi entropy can be computed with 
\begin{align}
    S_A^{(n)} = \frac{1}{1-n} \log \left\langle\prod_{i}\sigma_n(\tau_{a_i},x_{a_i})\tilde{\sigma}_n(\tau_{b_i},x_{b_i})\right\rangle_{\mathbb{K}^2}. 
\end{align}
The von Neumann entropy, or entanglement entropy, can be obtained by taking the $n\rightarrow1$ limit of the R\'enyi entropy 
\begin{align}
    S_A = \lim_{n\rightarrow1} S_A^{(n)}.
\end{align}

\begin{figure}[h]
    \centering
    \includegraphics[width=16cm]{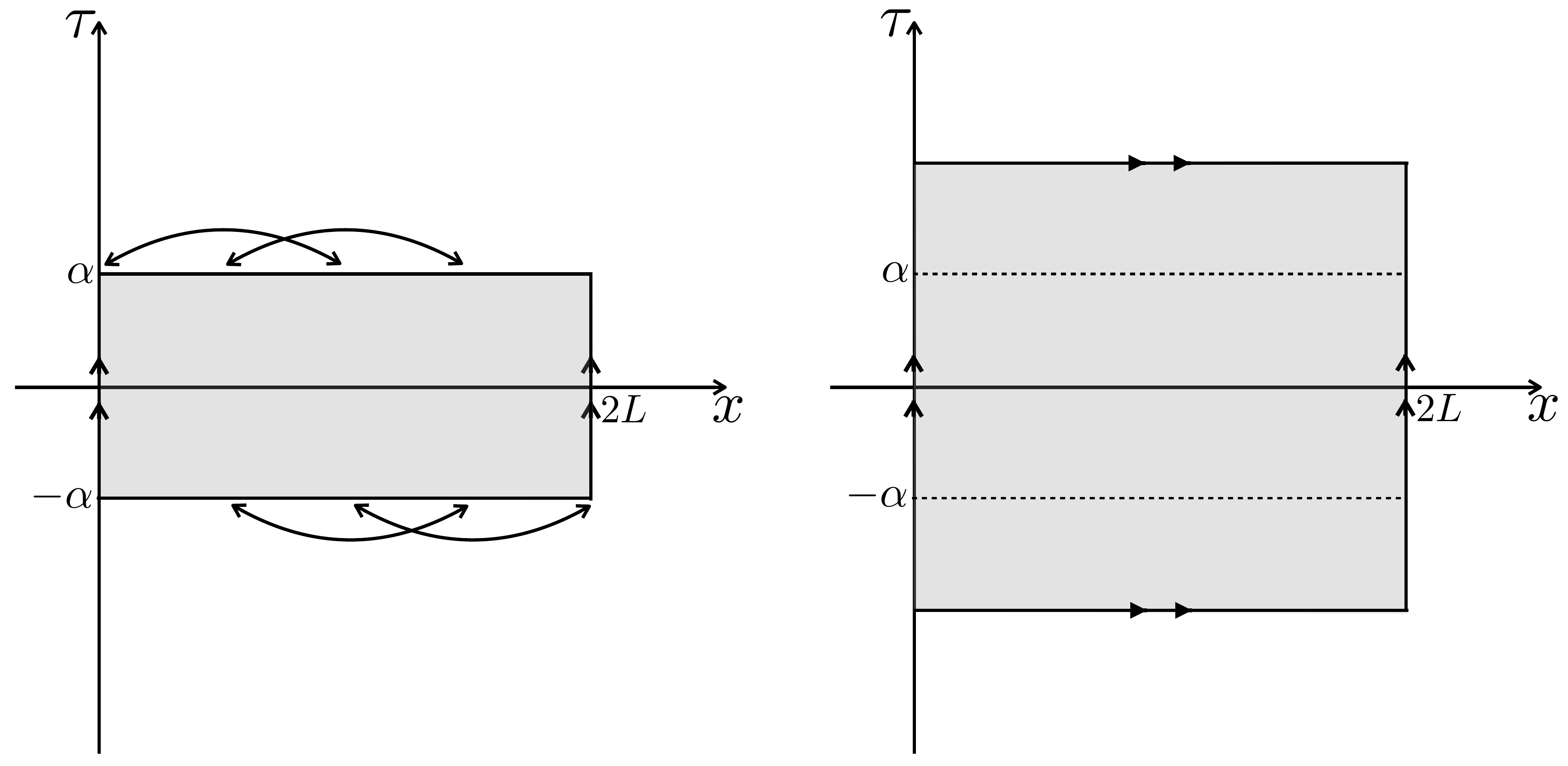}
    \caption{The left panel shows an alternative expression of the Klein bottle $\mathbb{K}^2$, and the right panel shows a torus $\mathbb{T}^2$. The $\mathbb{T}^2$ is a double of the $\mathbb{K}^2$, where the later can be obtained by performing the $\mathbb{Z}_2$ quotient $(\tau,x)\sim (2\alpha -\tau,x+L)$. At, e.g. the $\alpha/L \rightarrow0$ limit, correlation functions for operators at finite $x/L$ on the torus will approach to that on an infinite cylinder.}
    \label{fig:KandT}
\end{figure}

To evaluate the correlation function of twist operators on the Klein bottle, it is useful to apply the doubling trick. The Klein bottle parameterized by $\tau\in[-\alpha,\alpha]$ and $x\in[0,2L)$ admits a torus $\mathbb{T}^2$ parameterized by $\tau\in[-2\alpha,2\alpha)$ and $x\in[0,2L)$ as a double cover. The Klein bottle is obtained from the torus by performing the $\mathbb{Z}_2$ quotient $(\tau,x)\sim (2\alpha -\tau,x+L)$. 
Refer to Fig.~\ref{fig:KandT} for a visualization. 
Accordingly, inserting a twist operator $\sigma_n(\tau,x)$ on $\mathbb{K}^2$ has the same kinematics as simultaneously inserting $\sigma_n(\tau,x)$ and $\tilde{\sigma}_n(2\alpha-\tau,x+L)$ on $\mathbb{T}^2$ via the doubling trick. 

More precisely, 
since $\sigma_n(\tau,x)$ behaves as a scalar primary in the $n$-replicated CFT, we can decompose it into the left-moving (holomorphic) part and the right-moving (anti-holomorphic) part as 
\begin{align}
    \sigma_n(\tau,x) = \sigma_n^L(\tau,x)\sigma_n^R(\tau,x).
\end{align}
The correlation function of the twist operators on $\mathbb{K}^2$, 
\begin{align}\label{eq:K2point}   &\left\langle\prod_{i}\sigma_n(\tau_{a_i},x_{a_i})\tilde{\sigma}_n(\tau_{b_i},x_{b_i})\right\rangle_{\mathbb{K}^2} 
\end{align}
then has the same kinematics as the correlation function of the left-moving part of the twist operators on the torus $\mathbb{T}^2$, 
\begin{align}\label{eq:T4point} \left\langle\left(\prod_{i}\sigma_n^L(\tau_{a_i},x_{a_i})\tilde{\sigma}_n^L(\tau_{b_i},x_{b_i})\right) \left( \prod_{j}\tilde{\sigma}_n^L(2\alpha-\tau_{a_j},x_{a_j}+L){\sigma}_n^L(2\alpha-\tau_{b_j},x_{b_j}+L) \right) \right\rangle_{\mathbb{T}^2} 
\end{align}
Note that the mirror image of $\sigma_n$ turns out to be $\tilde{\sigma}_n$. 

The dynamics, i.e. OPE coefficients appearing in this correlation function,  depend on the details of the CFT and the choices of the crosscap condition. 
In the following, however, we study some universal behaviors of entanglement R\'enyi entropy in regularized crosscap states $e^{-\alpha H}|C\rangle$ and its time evolution in the quench dynamics. The strategy is to focus on situations where the correlation function can be reduced to the OPE limits of pairs of $\sigma_n^L$ and $\tilde{\sigma}_n^L$, such that the vacuum blocks dominate the computations and the CFT data dependence does not appear. 
We will not consider the OPE limits between $\sigma_n^L$ and $\sigma_n^L$ or those between $\tilde{\sigma}_n^L$ and $\tilde{\sigma}_n^L$, since the identity primary generically does not appear in these limits. We will leave the study of these OPE limits as future works. 

\subsection{Entanglement Structure of Crosscap States}

Following the strategy sketched above, let us study the entanglement structure of the regularized crosscap state $e^{-\alpha H}|C\rangle$, and see how it resembles that of the EAP states \eqref{eq:EAP_cutoff}. 

For the analysis below, it is useful to note that the two-point function of a scalar operator $O$ with $h_O=\bar{h}_O$ on an infinite cylinder parameterized by $x\in(-\infty,\infty)$ and $\tau \in [-2\alpha, 2\alpha)$ is 
\begin{align}           \braket{O(\tau_1,x_1)O(\tau_2,x_2)}_{\rm cylinder} = \left(\frac{\pi}{2\alpha}\right)^{4h_O} \left[2~\cosh \left(\frac{\pi(x_1-x_2)}{2\alpha}\right) - 2~\cos \left(\frac{\pi(\tau_1-\tau_2)}{2\alpha}\right)\right]^{-2h_O}, 
\end{align}
since the OPE limits of the Klein bottle correlators can be decomposed into different combinations of this cylinder two-point function. 

\paragraph{Single interval has volume law entanglement}~\par
The first situation we would like to consider is a single interval subsystem $A$ given by $0\leq x \leq l$. 
At the thermodynamic limit $\alpha/L \ll 1$, $l/L \ll 1$, 
\begin{align}
    &~~~\braket{\sigma_n(0,0)\tilde{\sigma}_n(0,l)}_{\mathbb{K}^2} \nonumber\\
    &\sim {\braket{\sigma_n^L(0,0)\tilde{\sigma}_n^L(0,l)\sigma_n^L(2\alpha,l+L)\tilde{\sigma}_n^L(2\alpha,L)}_{\mathbb{T}^2}} \nonumber\\
    &\sim \left[{\braket{\sigma_n(0,0)\tilde{\sigma}_n(0,l)}_{\rm cylinder}\braket{\sigma_n(2\alpha,l+L)\tilde{\sigma}_n(2\alpha,L)}_{\rm cylinder}}\right]^{1/2}  \nonumber\\
    &= \left(\frac{\pi}{2\alpha}\right)^{4h_n} \left[2~\cosh \left(\frac{\pi l}{2\alpha}\right)-2\right]^{-2h_n}
    = \left(\frac{\pi}{4\alpha}\right)^{4h_n} \left[~\sinh \left(\frac{\pi l}{4\alpha}\right)\right]^{-4h_n}.
\end{align}
In the second line, we have firstly used the fact that ${\braket{\sigma_n^L(0,0)\tilde{\sigma}_n^L(0,l)\sigma_n^L(2\alpha,l+L)\tilde{\sigma}_n^L(2\alpha,L)}_{\mathbb{T}^2}}$ and $\braket{\sigma_n(0,0)\tilde{\sigma}_n(0,l)}_{\mathbb{K}^2}$ have the same kinematics. More precisely, both of them can be decomposed into the form $\sum_{p}A_{\sigma_n \tilde{\sigma}_n p} F_p(l,\alpha,L)$ where $A_{\sigma_n \tilde{\sigma}_n p}$ is the data originating from the OPE coefficients and $F_p(l,\alpha,L)$ is the function originating from the conformal blocks. $p$ is a parameter labeling the primaries in the conformal block expansion.  $\braket{\sigma_n(0,0)\tilde{\sigma}_n(0,l)}_{\mathbb{K}^2}$ and ${\braket{\sigma_n^L(0,0)\tilde{\sigma}_n^L(0,l)\sigma_n^L(2\alpha,l+L)\tilde{\sigma}_n^L(2\alpha,L)}_{\mathbb{T}^2}}$  have the same $F_p(l,\alpha,L)$'s but different $A_{\sigma_n \tilde{\sigma}_n p}$'s. However, both of them are dominated by the vacuum block $p = \mathbb{I}$ and hence $\sim F_{\mathbb{I}}(l,\alpha,L)$
at the limit under consideration, which leads to the second line. The third line presents a convenient way to evaluate $F_{\mathbb{I}}(l,\alpha,L)$ again thanks to the thermodynamic limit, which leads to the final result in the last line. 

Accordingly, the entanglement R\'enyi entropy for the single interval subsystem $A$ reads, 
\begin{align}\label{eq:universal_single_interval}
    S_A^{(n)} 
    = \frac{c}{6}\left(1+\frac{1}{n}\right) \log \left[ \frac{4\alpha}{\pi \epsilon} \sinh \left(\frac{\pi l}{4\alpha}\right)\right], ~~~\left( \alpha/L \ll 1,~~l/L \ll 1 \right),
\end{align}
where $\epsilon$ is a UV cutoff corresponding to the lattice distance. Taking $n\rightarrow 1$, we obtain the entanglement entropy 
\begin{align}
    S_A 
    = \frac{c}{3} \log \left[ \frac{4\alpha}{\pi \epsilon} \sinh \left(\frac{\pi l}{4\alpha}\right)\right], ~~~\left( \alpha/L \ll 1,~~l/L \ll 1 \right). 
\end{align}
When the temperature is high, i.e. $l/\alpha \gg 1$, we have 
\begin{align}
     S_A^{(n)} = \frac{\pi c l}{24\alpha} \left(1+\frac{1}{n}\right) + \cdots, ~~~\left( \alpha/L \ll 1,~~l/L \ll 1, ~~\alpha/l \ll 1 \right),
\end{align}
and the subsystem $A$ exhibits the volume law entanglement. 

This is nothing but the thermal spectrum for a system at inverse temperature 
\begin{align}
    \beta = 4\alpha.
\end{align}
Therefore, a regularized crosscap state with cutoff $\alpha$ exhibits the same entanglement spectrum for single interval subsystems as the EAP state \eqref{eq:EAP_cutoff} at inverse temperature $\beta = 4\alpha$, at the thermodynamic limit.\footnote{It is worthwhile to emphasize that the effective inverse temperature $\beta$ is related to the Euclidean time evolution $\alpha$ as $\beta = 4\alpha$ rather than $\beta = 2\alpha$, which might appear more natural to some readers based on intuition. For the crosscap state, this relation is straightforward to see when the Klein bottle is viewed as a $\mathbb{Z}_2$ quotient of the torus. However, it is less straightforward in the case of the EAP states; see \cite{CY24,Yoneta24}. A similar phenomenon also appears when considering conformal boundary states with imaginary time evolution; see, e.g., \cite{CC05,HM13}. An effective temperature of the form $\beta = 2\alpha$ arises in other contexts, such as the minimally entangled typical thermal states \cite{White09} and the thermal pure quantum states \cite{SS11}. }  

Similarly, if we focus on a subsystem which is larger than a half of the whole system $l > L$ while the complement is finite at the thermodynamic limit $(2L-l)/L \ll 1$, the entanglement R\'enyi entropy reads 
\begin{align}\label{eq:universal_single_interval_2}
    S_A^{(n)} 
    = \frac{c}{6}\left(1+\frac{1}{n}\right) \log \left[ \frac{4\alpha}{\pi \epsilon} \sinh \left(\frac{\pi (2L-l)}{4\alpha}\right)\right], ~~~\left( \alpha/L \ll 1,~~(2L-l)/L \ll 1 \right),
\end{align}
manifesting the pure state nature of $e^{-\alpha H} \ket{C}$. 

\paragraph{Antipodal double intervals have area law entanglement}~\par
The next example we would like to consider is the double interval subsystem $A = \{x~|~0\leq x \leq l\} \cup \{x~|~L \leq x \leq L+l\}$, where the two components are located antipodally. We again take the thermodynamic limit $\alpha/L \ll 1$. 
In the regime $\alpha/l \ll 1$, the twist operator $\sigma_n(0,x)$ is very close to the mirror image of the twist operator $\sigma_n(0,L+x)$. Therefore, the 4-point function 
\begin{align}
    &~~~\braket{\sigma_n(0,0)\tilde{\sigma}_n(0,l)\sigma_n(0,L)\tilde{\sigma}_n(0,L+l)}_{\mathbb{K}^2} \nonumber\\
    &\sim{\braket{\sigma_n^L(0,0)\tilde{\sigma}_n^L(0,l)\sigma_n^L(0,L)\tilde{\sigma}_n^L(0,L+l)
    \tilde{\sigma}_n^L(2\alpha,0){\sigma}_n^L(2\alpha,l)\tilde{\sigma}_n^L(2\alpha,L){\sigma}_n^L(2\alpha,L+l)}_{\mathbb{T}^2}} \nonumber\\
    &\sim \braket{\sigma_n(0,0)\tilde{\sigma}_n(2\alpha,0)}_{\rm cylinder}^2 \nonumber \\
    &= \left(\frac{\pi}{4\alpha}\right)^{8h_n} .
\end{align}
Accordingly, the entanglement R\'enyi entropy for the double interval $A$
\begin{align}\label{eq:double_initial}
    S_A^{(n)} 
    = \frac{c}{3}\left(1+\frac{1}{n}\right) \log \left[ \frac{4\alpha}{\pi \epsilon} \right] , ~~~\left( \alpha/L \ll 1, ~~\alpha/l \ll 1 \right), 
\end{align}
which exhibits an area law entanglement. In particular, if we consider the high temperature limit and take $\alpha/\epsilon = O(1)$, then the antipodal double interval $A$ has almost no entanglement with its complement. This again matches the entanglment structure of antipodally located subsystems in EAP states. 

Based on the universal behavior of the entanglement R\'enyi entropy studied above, we can say that crosscap states serve as the analogue of EAP states in generic CFTs. This contrasts the conformal boundary states satisfying 
\begin{align}
    (L_n - \bar{L}_{-n})\ket{B} = 0, 
\end{align}
which essentially does not contain spatial entanglement at the leading order \cite{MRTW14}. 

\subsection{Time Evolution after the Crosscap Quench}\label{sec:universal_time_evolution}
Let us then study the time evolution of the entanglement entropy after the crosscap quench. In other words, we are going to focus on the state 
\begin{align}
    e^{-itH} e^{-\alpha H} |C\rangle. 
\end{align}
To this end, we compute the correlation function of twist operators for general insertion points $w=x+i\tau$, and then perform the analytic continuation $\tau\rightarrow it$. Some universal behaviors are investigated below. 

\paragraph{Single Interval is in equilibrium}~\par
For a single interval subsystem $A$ given by $0\leq x \leq l$. At the thermodynamic limit $\alpha/L \ll 1$, $l/L \ll 1$ and $t/L \ll 1$, 

\begin{align}
    \left.\braket{\sigma_n(\tau,0)\tilde{\sigma}_n(\tau,l)}_{\mathbb{K}^2}\right|_{\tau \rightarrow it} &\sim \left.\braket{\sigma_n(\tau,0)\tilde{\sigma}_n(\tau,l)\sigma_n(2\alpha-\tau,l+L)\tilde{\sigma}_n(2\alpha-\tau,L)}_{\mathbb{T}^2}^{1/2}\right|_{\tau\rightarrow it} \nonumber\\
    &\sim \braket{\sigma_n(0,-t)\tilde{\sigma}_n(0,l-t)\sigma_n(2\alpha,l+L+t)\tilde{\sigma}_n(2\alpha,L+t)}_{\mathbb{T}^2}^{1/2}\nonumber\\
    &\sim \braket{\sigma_n(0,-t)\tilde{\sigma}_n(0,l-t)}_{\rm cylinder}\braket{\sigma_n(2\alpha,l+\pi+t)\tilde{\sigma}_n(2\alpha,\pi+t)}_{\rm cylinder}^{1/2} \nonumber\\
    &= \left(\frac{\pi}{4\alpha}\right)^{4h_n} \left[~\sinh \left(\frac{\pi l}{4\alpha}\right)\right]^{-4h_n}
\end{align}
Therefore, the entanglement R\'enyi entropy is 
\begin{align}
    S_A^{(n)} 
    = \frac{c}{6}\left(1+\frac{1}{n}\right) \log \left[ \frac{4\alpha}{\pi \epsilon} \sinh \left(\frac{\pi l}{4\alpha}\right)\right] , ~~~\left( \alpha/L \ll 1,~~l/L \ll 1,~~t/L \ll 1 \right),
\end{align}
and the entanglement entropy is 
\begin{align}
    S_A
    = \frac{c}{3} \log \left[ \frac{4\alpha}{\pi \epsilon} \sinh \left(\frac{\pi l}{4\alpha}\right)\right] , ~~~\left( \alpha/L \ll 1,~~l/L \ll 1,~~t/L \ll 1 \right).
\end{align}
From this expression, we can see that, at the thermodynamic limit, the entanglement spectrum for a finite-size single interval does not change within finite time in the crosscap quench dynamics. In other words, such a single interval subsystem is in thermal equilibrium at the very beginning and stays in thermal equilibrium under the time evolution.

\paragraph{Antipodal double intervals evolve in a nontrivial way}~\par
Let us then consider the time evolution of the entanglement entropy 
for the double interval subsystem $A = \{x~|~0\leq x \leq l\} \cup \{x~|~L \leq x \leq L+l\}$ located antipodally. In a similar manner, at $\alpha/L \ll 1, \alpha/l \ll 1, t/l \ll 1$,
\begin{align}
    &~~~\left.\braket{\sigma_n(\tau,0)\tilde{\sigma}_n(\tau,l)\sigma_n(\tau,L)\tilde{\sigma}_n(\tau,L+l)}_{\mathbb{K}^2}\right|_{\tau\rightarrow it} \nonumber\\
    &\sim \braket{\sigma_n(0,-t)\tilde{\sigma}_n(0,l-t)\sigma_n(0,L-t)\tilde{\sigma}_n(0,L+l-t)
    \tilde{\sigma}_n(2\alpha,t){\sigma}_n(2\alpha,l+t)\tilde{\sigma}_n(2\alpha,L+t){\sigma}_n(2\alpha,L+l+t)}_{\mathbb{T}^2}^{1/2} \nonumber\\
    &\sim \braket{\sigma_n(0,-t)\tilde{\sigma}_n(2\alpha,t)}_{\rm cylinder}^2 \nonumber\\
    &= \left(\frac{\pi}{4\alpha}\right)^{8h_n} \cosh \left(\frac{\pi t}{2\alpha}\right)^{-8h_n}
\end{align}
The entanglement R\'enyi entropy then turns out to be 
\begin{align}
    S_A^{(n)} = 
    \frac{c}{3}\left(1+\frac{1}{n}\right) \log \left[ \frac{4\alpha}{\pi \epsilon} \cosh \left(\frac{\pi t}{2\alpha}\right) \right] , ~~~\left( \alpha/L \ll 1, ~~\alpha/l \ll 1,~~t/l \ll 1 \right),
\end{align}
which exhibits a linear growth at late time 
\begin{align}
     S_A^{(n)} = \frac{\pi c t}{6\alpha} \left(1+\frac{1}{n}\right) + \cdots, ~~~\left( \alpha/L \ll 1,~~l/L \ll 1, ,~~t/l \ll 1~~\alpha/t \ll 1 \right). 
\end{align}
If we instead focus on the case $l/L\ll 1$, and look at the behavior at very late time $l/t \ll 1$ (but suppressed at the thermodynamic limit $t/L \ll 1$), then we can see that the entanglement entropy saturates and becomes identical to that of the thermal state, 
\begin{align}
    S_A^{(n)} = 
    \frac{c}{3}\left(1+\frac{1}{n}\right) \log \left[ \frac{4\alpha}{\pi \epsilon} \sinh \left(\frac{\pi l}{4\alpha}\right)\right], ~~~\left( \alpha/L \ll 1, ~~l/L \ll 1,~~l/t \ll 1, ~~ t/L \ll 1 \right). 
\end{align}
By comparing these expressions with that at the initial state \eqref{eq:double_initial}, we can see that, under the crosscap quench, such a antipodal double interval subsystem initially exhibits the area law entanglement, then experiences a linear growth of entanglement, and eventually converges to the volume law thermal spectrum. This dynamics shows how the initial crosscap state, which is already in thermal equilibrium but engineered with highly organized entanglement structure, relaxes under time evolution into a more scrambled few-body thermal state. Here, by a few-body thermal state we mean a state that is indistinguishable from a Gibbs state not only with respect to local observables, but also with respect to few-body, yet geometrically nonlocal, observables~\cite{MIKU17}.

Another notable point is that the entanglement time evolution of antipodally located double intervals under the crosscap quench is qualitatively similar to that of a single interval under the boundary state quench starting from $e^{-\alpha H}|B\rangle$ \cite{CC05,HM13,Ugajin13}. This can be easily interpreted using the EAP state \eqref{eq:EAP} analogue in the following way. An antipodally located double interval subsystem in the EAP state consisting of 2$N$ spins can be equivalently regarded as a single interval subsystem in a $N$-site circle where each site has two spins. Such an $N$-site state is a product state in terms of spatial separation, which can be regarded as an analogue of the conformal boundary state \cite{MRTW14}.

\section{Holographic Crosscap Quench}\label{sec:HolographiCQ}

In the previous section, we formulated the crosscap quench in CFT and studied some universal behaviors of the entanglement R\'enyi entropy. 

In this section, we study the crosscap quench in the AdS/CFT correspondence \cite{Maldacena01,GKP98,Witten98} and apply the Ryu-Takayanagi (RT) formula \cite{RT06,RT06b} and its covariant version, the Hubeny-Rangamani-Takayanagi (HRT) formula \cite{HRT07}, to study the entanglement entropy in the holographic crosscap quench dynamics, assuming their validity. This serves as an explicit example of a chaotic critical system, where the entanglement entropy can be explicitly computed. 

As we have already seen in the previous section, the crosscap quench is described by the unitary time evolution starting from a regularized crosscap state
\begin{align}
    e^{-it H} e^{-\alpha H } \ket{C}, 
\end{align}
which is prepared by a Schwinger-Keldysh path integral from a Klein bottle. The gravity dual of such a CFT has been worked out in \cite{Maldacena01} and turned out to be the so-called $\mathbb{RP}^2$ geon geometry in AdS$_3$ \cite{LM98}. In addition, the holographic entanglement entropy associated with the geon geometry has been investigated in \cite{Maxfield14}. Therefore, the computations presented in this section are not new, but are to be reinterpreted as the crosscap quench in holography and compared with results in other sections. 

In the following, we first present the AdS dual of such a crosscap quench, and then compute the holographic entanglement entropy. To match the standard convention used in the gravity side, we set 
\begin{align}
    \alpha = \frac{\beta}{4}, ~~~2L = 2\pi,
\end{align}
throughout this section. 

\subsection{The Gravity Dual}
In the AdS$_3$/CFT$_2$ correspondence, when the boundary geometry is a Klein bottle $\mathbb{K}^2$ with a sufficiently high temperature (i.e. sufficiently small $\beta$), the bulk dual is given by the Euclidean geon geometry \cite{Maldacena01,GR14,MR16,Wei24}. The metric is identical to that of a BTZ black hole: 
\begin{align}
    ds^2 = \left(r^2 - r_H^2\right) d\tau^2 +  \frac{dr^2}{r^2 - r_H^2} + r^2 dx^2, 
\end{align}
where $r = r_H \equiv 2\pi/\beta$ is the location of the event horizon. The difference from the BTZ black hole, however, is a $\mathbb{Z}_2$ quotient: 
\begin{align}
    \left(\tau, x ,r\right) \sim \left(\frac{\beta}{2}-\tau, x+\pi ,r\right). 
\end{align} 
Let us firstly focues on the geometry of the $\tau = 0$ time slice. While the $r>r_H$ part of the geoemtry is identical to that of the BTZ black hole, a crosscap is inserted at the horizon $r = r_H$. As the result, the topology of the $\tau = 0$ slice is a M\"obius strip. This is the gravity dual of $e^{-\frac{\beta}{4}H}|C\rangle$, the initial state of the crosscap quench. 

Performing the analytic continuation $\tau = i t$, the metric turns out to be 
\begin{align}
    ds^2 = -\left(r^2 - r_H^2\right) dt^2 +  \frac{dr^2}{r^2 - r_H^2} + r^2 dx^2. 
\end{align}
To describe the interior of the black hole, let us extend the coordinate to the Kruskal coordinate with 
\begin{align}
    & \tilde{U} \equiv \pm \sqrt{\frac{r-r_H}{r+r_H}} e^{r_H t_{\mp}}, \\
    & \tilde{V} \equiv \mp \sqrt{\frac{r-r_H}{r+r_H}} e^{-r_H t_{\mp}},
\end{align}
where $t_-$ is the time $t$ obtained by $\tau = it_-$, and $t_+$ is the time direction on the other side of the black hole obtained by $\tau = \beta/2-it_+$. With the Kruskal coordinate, the metric then turns out to be 
\begin{align}
    ds^2 = \frac{-4~ d\tilde{U}d\tilde{V} + (-1 + \tilde{U}\tilde{V})^2 ~r_H^2 ~dx^2}{\left(1+\tilde{U}\tilde{V}\right)^2}. 
\end{align}
We can also introduce a time-like coordinate $\tilde{T}$ and a space-like coordinate $\tilde{X}$ with 
\begin{align}
    & \tilde{U} \equiv \tilde{T} - \tilde{X}, \\
    & \tilde{V} \equiv \tilde{T} + \tilde{X},
\end{align}
with which the metric is now
\begin{align}
    ds^2 = \frac{-4~ d\tilde{T}^2 + 4~ d\tilde{X}^2 + (-1 + \tilde{T}^2-\tilde{X}^2)^2 ~r_H^2 ~dx^2}{\left(1+\tilde{T}^2-\tilde{X}^2\right)^2}. 
\end{align}
The $\mathbb{Z}^2$ quotient is 
\begin{align}
    \left(\tilde{T}, x, \tilde{X}\right) \sim \left(\tilde{T}, x+\pi, -\tilde{X}\right). 
\end{align}
This one-sided black hole is the gravity dual of the crosscap quench dynamics $e^{-itH}e^{-\frac{\beta}{4}H}|C\rangle$. The topology of each Cauchy slice is the M\"obius strip and hence non-orientable. However, this non-orientability cannot be verified by an asymptotic observer. This is because that the crosscap sits at $\tilde{X}=0$, which is behind the event horizon and hence can never be probed by any communication with an asymptotic observer. 

\subsection{Holographic Entanglement Entropy and Homology Condition}
In the following, we apply the Ryu-Takayanagi (RT) formula \cite{RT06,RT06b} and the Hubeny-Rangamani-Takayanagi (HRT) formula \cite{HRT07} to compute the entanglement entropy in the crosscap quench dynamics. 

In AdS$_{d+1}$/CFT$_d$, a time slice of the CFT describes a quantum state of a $(d-1)$-dimensional quantum system. After taking a time slice, we divide the spatial region of the CFT into two subregions $A$ and $B$. The entanglement entropy of subsystem $A$, denoted as $S_A$, can then be computed using the HRT formula. The procedure is as follows. The subregion $A$ is $(d-1)$-dimensional and hence codimension-2 from the AdS point of view. We are going to look for a spacelike codimension-2 surface $\gamma_A$, which shares the boundary with the subregion $A$, i.e. $\partial \gamma_A = \partial A$, satisfies the homology constraint, 
\begin{align}\label{eq:homology}
    &\gamma_A \sim A ~~ \xLeftrightarrow[]{\rm def.} ~~\exists~\text{spacelike codimension-1 $\mathcal{R}_A$ surface s.t. $\partial{\mathcal{R}_A} = \gamma_A \cup A$}, 
\end{align}
and is extremal under small perturbations. 
Among all such codimension-2 surface $\gamma_A$, the one with the minimal area (called the HRT surface) is related to the entanglement entropy of subsystem $A$ as, 
\begin{align}
    S_A = \min \underset{\substack{\partial \gamma_A = \partial A \\ \gamma_A \sim A}}{{\rm ext}} \frac{{\text{Area}(\gamma_A)}}{4 G_N^{(d+1)}}, 
\end{align}
where $G_N^{(d+1)}$ is the Newton constant of the bulk AdS. In the following, we will denote the HRT surface as $\Gamma_A$ to make a distinction with other candidates. The HRT formula is applicable in setups with general time dependence. In special cases when the whole setup is static or when the time slice we are focusing on posits a time reflection symmetry, the HRT formula reduces to the RT formula, where one first restricts to the time-reflection symmetric Cauchy slice, and then looks for the minimal surface in it. 

In the following, we would like to compute the entanglement entropy of the crosscap quench in holography. Since we are interested in the time evolution, we need to apply the HRT formula. However, the $\tau = 0$ slice is symmetric under the time-reflection, so the RT formula is applicable there. 

Moreover, each Cauchy slice of the bulk dual of the crosscap quench includes a crosscap in it, and is hence non-orientable. Therefore, the homology constraint \eqref{eq:homology} will turn out to be crucial in the holographic analysis. 

We will firstly focus on the $t = 0$ time slice, where the RT formula is applicable to illustrate the nontrivialness of the homology condition, and then compute the time evolution of the entanglement entropy. 

\subsection{EE in the Initial State} 
Let us firstly apply the RT formula to study the entanglement entropy associated to the initial state at $t=0$. In this process, the treatment of the homology condition will become clear. 

\paragraph{The whole system}~\par
\begin{figure}
    \centering
    \includegraphics[width=16cm]{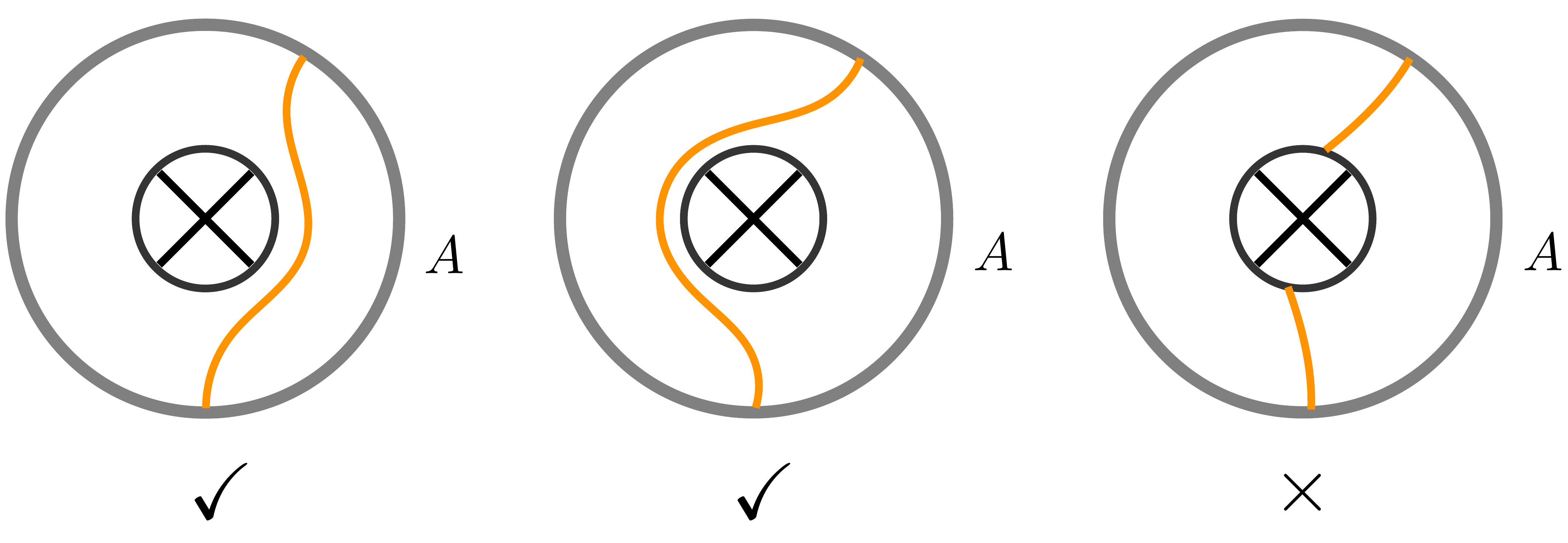}
    \caption{Three types of $\gamma_A$ which satisfies $\partial \gamma_A \sim \partial A$, where $A$ is a single interval. The left two satisfies the homology condition defined in \eqref{eq:homology}, while the rightmost does not.}
    \label{fig:RT_single_interval}
\end{figure}

The simplest choice of subsystem $A$ is the entire system. In this case, the state $\rho_A$ is a pure state and hence the entanglement entropy must be zero. Let us confirm this results is reproduced by the RT formula. The topology of the $t=0$ time slice is a M\"obius strip as shown in Fig.~\ref{fig:RT_single_interval}. It is simple but essential to note that the empty set $\emptyset$ satisfies the homology constraint  \eqref{eq:homology} by identifying the whole $t=0$ time slice as $\mathcal{R}_A$. Therefore, the holographic EE computed by the RT formula is indeed zero. 

\paragraph{Single Interval}~\par
Let us then take the subsystem $A$ to be a single interval $0\leq x \leq l$. In this case, there are two types of geodesics $\gamma_A$ such that $\partial \gamma_A = \partial A$ and satisfy the homology constraint $\gamma_A \sim A$. In the first case, the geodesic $\gamma_A$ can be obtained by continuously deforming $A$, and the corresponding $\mathcal{R}_A$ does not contain a crosscap. In the second case, the geodesic $\gamma_A$ cannot be obtained by continuously deforming $A$, and the corresponding $\mathcal{R}_A$ contains the crosscap. There also exist geodesics that do not satisfy the homology constraint and therefore do not serve as the candidates of RT surfaces. This happens when the geodesic $\gamma_A$ ``goes through'' the crosscap, and cannot be obtained from $A$ by a continuous deformation. See
Fig.~\ref{fig:RT_single_interval} for a sketch. 
As a result, there are two competing candidates of the RT surfaces, and the EE turns out to be 
\begin{equation} \label{eq:single_interval}
S_A=
\min \begin{cases}
\frac{c}{3} \log\left[\frac{\beta}{\pi\epsilon} \,\sinh \left(\frac{\pi l}{\beta} \right) \right]\\
\frac{c}{3} \log\left[\frac{\beta}{\pi\epsilon} \,\sinh \left(\frac{\pi (2\pi-l)}{\beta} \right) \right]. 
\end{cases} 
\end{equation}
Here, we have used the Brown-Henneaux relation $1/4G_N = c/6$ \cite{BH86}. 
We can also check that this result matches the universal behavior \eqref{eq:universal_single_interval} and \eqref{eq:universal_single_interval_2} computed in the previous section.

\paragraph{Antipodal double interval}~\par
\begin{figure}
    \centering
    \includegraphics[width=16cm]{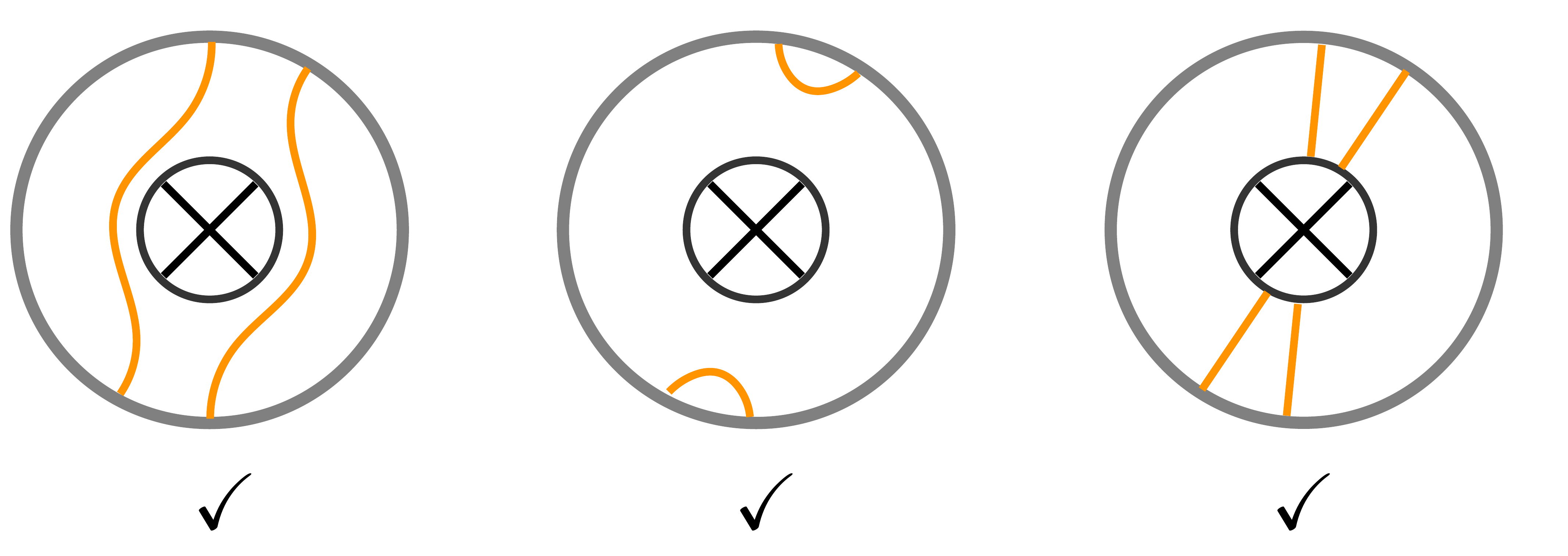}
    \caption{Three types of $\gamma_A$ which satisfies the homology condition defined in \eqref{eq:homology} when $A$ is antipodally located double intervals.}
    \label{fig:RT_double_interval}
\end{figure}

The next example we would like to consider is the antipodally located double interval $0\leq x \leq l$ and $\pi \leq x \leq \pi + l$. In this case, there are three candidates of RT surfaces satisfying the homology constraint as shown in Fig.~\ref{fig:RT_double_interval}. Compared to the single interval case, the RT surface now can ``go across'' the crosscap. As a result, the EE turns out to be a competition between the three candidates. 
\begin{equation}
S_A=
\min \begin{cases}
\frac{2c}{3} \log\left[\frac{\beta}{\pi\epsilon} \,\sinh \left(\frac{\pi l}{\beta} \right) \right]\\
\frac{2c}{3} \log\left[\frac{\beta}{\pi\epsilon} \right]\\
\frac{2c}{3} \log\left[\frac{\beta}{\pi\epsilon} \,\sinh \left(\frac{\pi (2\pi-l)}{\beta} \right) \right]. 
\end{cases}
\end{equation}
This is again consistent with the universal behavior \eqref{eq:double_initial} investigated on the CFT side. 
We can see that the EE firstly grow, and then saturate, and then decrease. In other words, the crosscap state exhibits an area law for antipodally located double intervals. This is natural since the crosscap state $\ket{C}$ defined on $S^1$ can be regarded as a product state defined on $S^1/Z_2$ where the $Z_2$ is the antipodal identification map, and it is known that the Euclidean time evolution of a product state in 1D quantum many-body system generically exhibits an EE sclaes as $O(c l^0 \log \beta )$ \cite{KTWY23}. 

\subsection{Time Evolution of EE after the Crosscap Quench}

Then we would like to compute the holographic EE under the time evolution. Similar to the reasoning in the $t=0$ case, the HRT surface for the entire system is the empty set $\Gamma_A = \emptyset$ and its EE stays to be zero. This manifests the unitarity of the time evolution. 

For the single interval, there are again two candidates of the RT surfaces. Both of them stay outside the event horizon and their lengths do not change in time. As a result, the EE for a single interval stays to be \eqref{eq:single_interval}. 

For the antipodal double intervals, on the other hand, the RT surface going across the crosscap travels inside the event horizon and exhibits a nontrivial time evolution, while the two other candidate RTs stay to be invariant under the time evolution. As a result, we have 
\begin{equation}
S_A=
\min \begin{cases}
\frac{2c}{3} \log\left[\frac{\beta}{\pi\epsilon} \,\sinh \left(\frac{\pi l}{\beta} \right) \right]\\
\frac{2c}{3} \log\left[\frac{\beta}{\pi\epsilon} \cosh \left(\frac{2\pi}{\beta}t\right)\right] \\
\frac{2c}{3} \log\left[\frac{\beta}{\pi\epsilon} \,\sinh \left(\frac{\pi (2\pi-l)}{\beta} \right) \right]. 
\end{cases}
\end{equation}
We can see that the EE exhibits a linear growth at late time, and eventually saturates to a volume law entanglement. These are also consistent with the universal behavior in CFT investigated in section \ref{sec:universal_time_evolution}.

\section{Crosscap Quench in Nonintegrable Spin Systems}\label{sec:chaotic_spin_system}
Returning to the lattice theory, we numerically compute the time evolution of entanglement under dynamics starting from an EAP state and compare the results with the field-theoretic predictions discussed in the previous sections.

We begin our numerical analysis with the case where the Hamiltonian is nonintegrable. Here, we consider the spin-$1/2$ Heisenberg chain with next-nearest-neighbor interactions, defined by the Hamiltonian
\begin{align}
    H = \sum_{j=1}^{N} \sigma_{j}^x \sigma_{j+1}^x + \sigma_{j}^y \sigma_{j+1}^y + \sigma_{j}^z \sigma_{j+1}^z
    + J \sum_{j=1}^{N} \sigma_{j}^x \sigma_{j+2}^x + \sigma_{j}^y \sigma_{j+2}^y + \sigma_{j}^z \sigma_{j+2}^z.
    \label{eq:NNNHeisenberg}
\end{align}
It was shown that this model is nonintegrable \cite{Shiraishi2024,Hsu1993}.
We set the coupling $J$ at the quantum critical point $J = 0.241167$, which is well described by a CFT with central charge $c = 1$ \cite{ON92,NO94,Eggert96}.

As an initial state, we choose the EAP state $\ket{\rm EAP}$ defined by Eq.~\eqref{eq:EAP} with $\ket{\Phi}_{j,j+N} \propto \ket{0}_{j}\ket{0}_{j+N}+\ket{1}_{j}\ket{1}_{j+N}$
\begin{align}
    \ket{\rm EAP} = \bigotimes_{j=1}^{N} \frac{\ket{0}_{j}\ket{0}_{j+N}+\ket{1}_{j}\ket{1}_{j+N}}{\sqrt{2}}.
    \label{eq:EAP_00}
\end{align}
Here, $\ket{0}_{j}$ and $\ket{1}_{j}$ are the eigenvectors of $\sigma_{j}^z$ in the local Hilbert space associated with site $j$.
For this choice of the EAP state, as shown in Ref.~\cite{Yoneta24}, the imaginary-time evolved EAP state $\ket{{\rm EAP}(\beta)} \propto e^{- \frac{\beta}{4} H} \ket{\rm EAP}$ is a thermal pure state at inverse temperature $\beta$ because the Hamiltonian \eqref{eq:NNNHeisenberg} is real in the spin basis. For simplicity, we set $\beta = 0$ in the following discussion. In this case, the entanglement entropy for a single interval and an antipodally located double interval, each with volume $l$, are exactly $\min\{l, 2L-l\} \log 2$ and $0$, respectively, simplifying the comparison between the pre- and post-quench states. Then the time-evolved state at time $t$ is given by
\begin{align}
    \ket{{\rm EAP}(t)} = e^{- i H t} \ket{\rm EAP}.
\end{align}
We numerically calculate the entanglement entropy $S_A$ of this state for various choices of subsystems $A$.

\begin{figure}[h]
    \centering
    \includegraphics[width=0.48\linewidth]{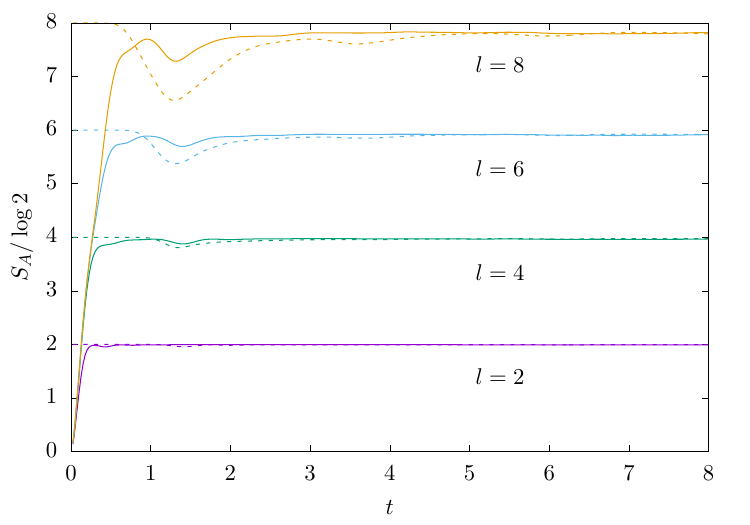}
    \caption{Entanglement entropy $S_A$ of the time-evolved EAP state $\ket{{\rm EAP}(t)}$ in the critical Heisenberg chain with next-nearest-neighbor interactions as a function of $t$ for total system size $2N=24$ and various subsystem size $l$, with solid lines representing the antipodally located double intervals $A = \{1, 2, \dots, l/2\} \cup \{N+1, N+2, \dots, N+l/2\}$ and dashed lines represent the single intervals $A = \{1, 2, \dots, l\}$.}
    \label{fig:NNNHeisenberg_L20_t}
\end{figure}

Figure~\ref{fig:NNNHeisenberg_L20_t} shows the dynamics of entanglement entropy for the single interval $A = \{1, 2, \dots, l\}$ and the antipodally located double interval $A = \{1, 2, \dots, l/2\} \cup \{N+1, N+2, \dots, N+l/2\}$. For {$l \ll L$}, the entanglement entropy exhibits an initial linear growth and saturates at a constant value, in well agreement with CFT predictions. Moreover, this saturated value coincides with the thermodynamic entropy at $\beta = 0$, $l \log 2$. These results implies that the EAP state acquires a complex correlation structure and relaxes into a few-body thermal state, which is indistinguishable from the Gibbs state not only with respect to local observables but also with respect to few-body, yet geometrically nonlocal, observables.

{
To characterize the long-time behavior, we introduce the time-averaged
entanglement entropy
\begin{align}
    \overline{S}_A = \frac{1}{t_2-t_1}\int_{t_1}^{t_2} dt S_A(t),
    \label{eq:timeavg_entropy}
\end{align}
where in our numerics we take $(t_1,t_2)=(8,32)$. We evaluate $\overline{S}_A$ by sampling $S_A(t)$ at uniform time steps $\Delta t = 0.025$ and taking the discrete average over these time points.
}

\begin{figure}[h]
    \begin{minipage}{0.48\linewidth}
        \centering
        \includegraphics[width=\linewidth]{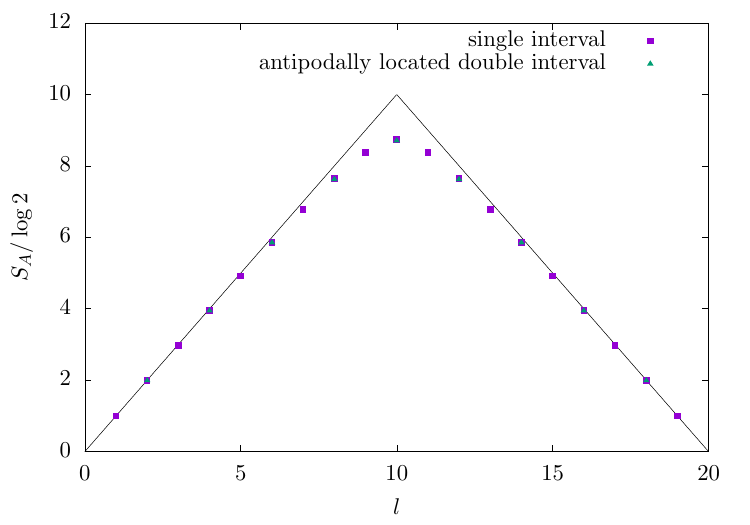}
        \subcaption{}
        \label{fig:NNNHeisenberg_L20_l-SA}
    \end{minipage}
    \begin{minipage}{0.48\linewidth}
        \centering
        \includegraphics[width=\linewidth]{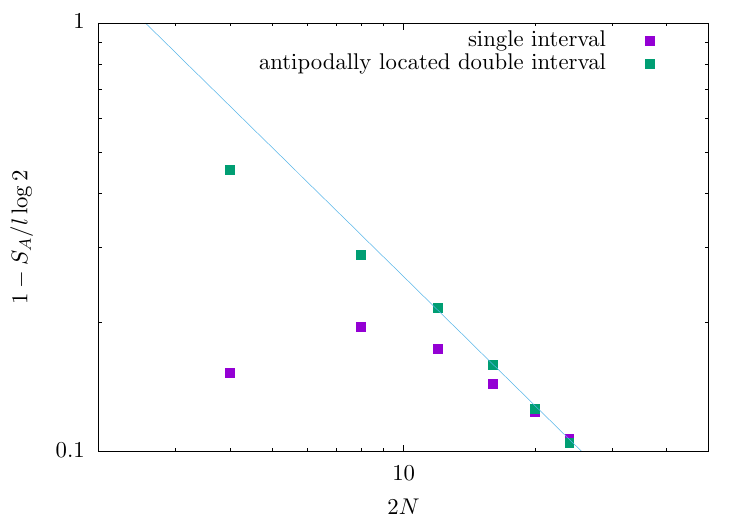}
        \subcaption{}
        \label{fig:NNNHeisenberg_L-dSA}
    \end{minipage}
    \caption{%
        (\subref{fig:NNNHeisenberg_L20_l-SA}) Time-averaged entanglement entropy of the time-evolved EAP state $\ket{{\rm EAP}(t)}$ in the critical Heisenberg chain with next-nearest-neighbor interactions, {averaged over the time interval $t \in [8, 32]$ with a uniform time step $\Delta t=0.025$}, as a function of the subsystem size $l$ for the single interval and the antipodally located double interval in a system of size $2N=20$. The solid line represents the entanglement entropy of a single interval in the initial EAP state, which exhibits maximal entanglement for any such interval.
        (\subref{fig:NNNHeisenberg_L-dSA}) Deviation from the maximum value at $l=N$. The solid line serves as a guide to the eye and is proportional to $N^{-1}$.
    }
\end{figure}

As illustrated in Fig.~\ref{fig:NNNHeisenberg_L20_l-SA}, the time-evolved EAP state $\ket{{\rm EAP}(t)}$ contains a volume-law entanglement with a slight deviation increasing as the subsystem size grows, and the deviation becomes most pronounced when the subsystem size is exactly half of the total system size. Then we plot the system-size dependence of the deviation in Fig.~\ref{fig:NNNHeisenberg_L-dSA}. It can be seen that the deviation is subextensive. This is consistent with the results obtained from the analysis of holographic CFTs. Furthermore, we find that the deviation scales as $N^0$ in the thermodynamic limit.


\section{Crosscap Quench in Integrable Spin Systems}\label{sec:integrable_spin_system}

In this section, for comparison with the nonintegrable case analyzed in the previous section, we examine the entanglement dynamics of the EAP state in an integrable critical quantum spin chain. Specifically, we consider the transverse-field Ising chain, which is integrable and exhibits quantum criticality, defined by the Hamiltonian
\begin{align}
    H = - \sum_{j=1}^{N} \sigma_{j}^z \sigma_{j+1}^z - g \sum_{j=1}^{N} \sigma_{j}^x.
\end{align}
The quantum critical point is located at $g = 1$ and is well described by CFT with central charge $c = 1/2$ \cite{Sachdev11,CC04}.
{This Hamiltonian is also real with respect to the spin basis.  Therefore,} as in the previous section, we consider the dynamics starting from the EAP state \eqref{eq:EAP_00} defined with $\ket{\Phi}_{j,j+N} \propto \ket{0}_{j}\ket{0}_{j+N} + \ket{1}_{j}\ket{1}_{j+N}$.

\begin{figure}[ht]
    \centering
    \includegraphics[width=0.48\linewidth]{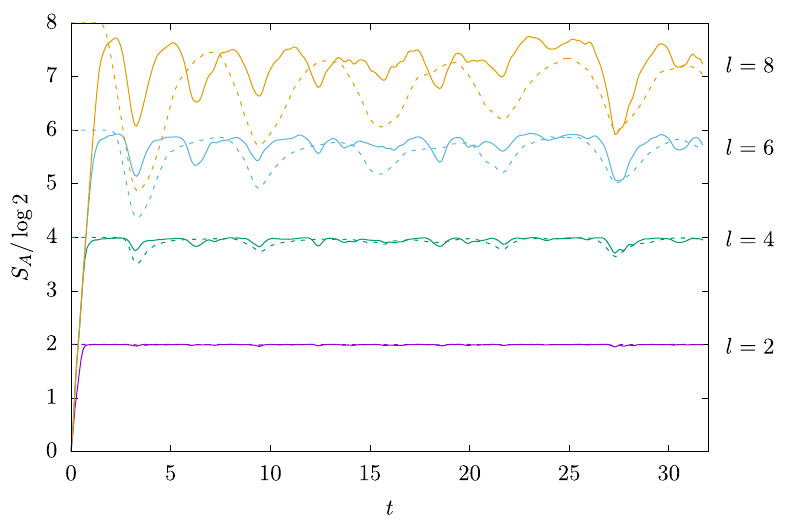}
    \caption{Entanglement entropy $S_A$ of the time-evolved EAP state $\ket{{\rm EAP}(t)}$ in the critical transverse-field Ising chain as a function of $t$ for total system size $2N=24$ and various subsystem size $l$, with solid lines representing the antipodally located double intervals $A = \{1, 2, \dots, l/2\} \cup \{N+1, N+2, \dots, N+l/2\}$ and dashed lines represent the single intervals $A = \{1, 2, \dots, l\}$.}
    \label{fig:TFIM_L24_t-SA}
\end{figure}

In Fig.~\ref{fig:TFIM_L24_t-SA}, we show the dynamics of the entanglement entropy for the single interval $A = \{1, 2, \dots, l\}$ and the antipodally located double interval $A = \{1, 2, \dots, l/2\} \cup \{N+1, N+2, \dots, N+l/2\}$. For $l \ll L$, similar to the nonintegrable case, the entanglement entropy for the antipodal intervals exhibits an initial linear growth followed by saturation at the thermodynamic entropy for $\beta = 0$. This reflects the universality of the result obtained from the CFT analysis in Section~\ref{sec:CQinCFT}, which { holds for the regime with a separation of scales $a \ll l \ll L$} independent of the details of the theory, such as the integrability. Consequently, even in the integrable systems, the EAP state relaxes into a few-body thermal state.

\begin{figure}[H]
    \begin{minipage}{0.48\linewidth}
        \centering
        \includegraphics[width=\linewidth]{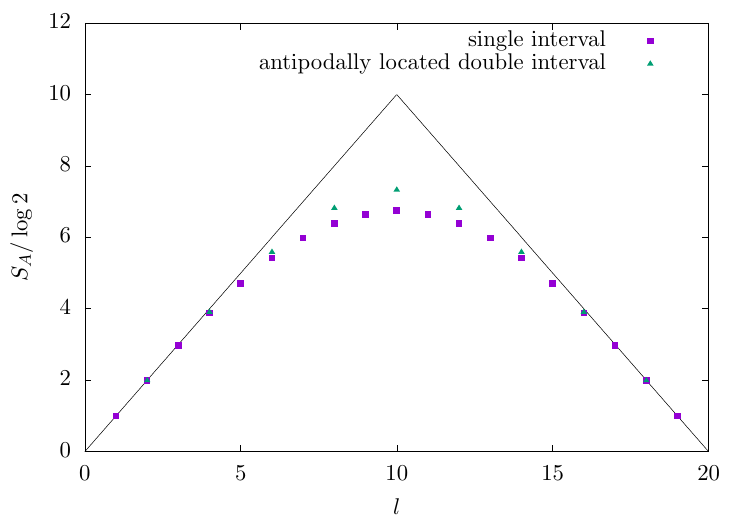}
        \subcaption{}
        \label{fig:TFIM_L20_l-SA}
    \end{minipage}
    \begin{minipage}{0.48\linewidth}
        \centering
        \includegraphics[width=\linewidth]{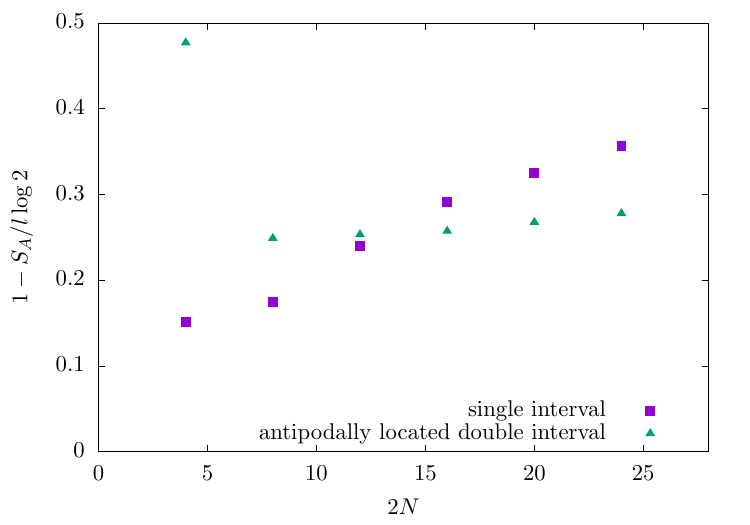}
        \subcaption{}
        \label{fig:TFIM_L-dSA}
    \end{minipage}
    \caption{%
        (\subref{fig:TFIM_L20_l-SA}) Time-averaged entanglement entropy of the time-evolved EAP state $\ket{{\rm EAP}(t)}$ in the critical transverse-field Ising chain, { averaged over the time interval $t \in [8, 32]$ with a uniform time step $\Delta t=0.025$}, as a function of the subsystem size $l$ for the single interval and the antipodally located double interval in a system of size $2N=20$. The solid line represents the entanglement entropy of a single interval in the initial EAP state, which exhibits maximal entanglement for any such interval.
        (\subref{fig:TFIM_L-dSA}) Deviation from the maximum value at $l=N$.
    }
\end{figure}

In contrast, when the subsystem size $l$ becomes comparable to the total system size, $l \sim N$, the situation differs. It can be seen that for large subsystems, the entanglement entropy does not equilibrate but instead oscillates. Furthermore, as shown in Figs.~\ref{fig:TFIM_L20_l-SA} and \ref{fig:TFIM_L-dSA}, the deviation from the maximum entanglement of the single interval initially contained in the EAP state, caused by time evolution, is extensive, unlike in nonintegrable systems.

Therefore, while the EAP state in an integrable system also relaxes into a few-body thermal state, the resulting state exhibits an entanglement structure that is qualitatively distinct from that of the nonintegrable cases.

\section{Conclusion}

In this paper, we have studied quantum quenches in 1D many-body systems starting from crosscap states and EAP states. While EAP states are originally defined in spin systems, we have argued that crosscap states serve as good analogues of them in CFT because they exhibit similar entanglement structures. This was demonstrated by studying the universal behavior of the entanglement Rényi entropy of CFT crosscap states. Based on this, we termed the quenches starting from crosscap states and EAP states as ``crosscap quenches.'' We then investigated the time evolution of entanglement in both general CFT and holographic settings of crosscap quenches, with the latter serving as an analytically tractable example of chaotic systems. Additionally, we numerically studied crosscap quenches in both a nonintegrable critical spin chain and an integrable critical spin chain.

In the nonintegrable systems studied in this paper, local subsystems, such as small single intervals, exhibit a volume law for entanglement and are expected to remain thermal throughout the dynamics. On the other hand, the entanglement entropy of antipodally located double intervals initially exhibits an area law, followed by linear growth, and eventually saturates to a volume law. This behavior illustrates how the initially well-organized thermal state becomes scrambled under time evolution and transitions into a state of few-body thermalization. Conversely, the integrable system deviates from the initially well-organized thermal equilibrium and exhibits oscillations in the time evolution of entanglement.

\section*{Acknowledgements}
We are grateful to Yuya Kusuki, Diandian Wang and Zhencheng Wang for useful discussions and
comments. 
ZW is supported by the Society of Fellows at Harvard University. 
YY is supported by the Special Postdoctoral Researchers Program at RIKEN.

\appendix

\bibliographystyle{jhep}
\bibliography{CrosscapQuench}

\end{document}